\documentclass{article}
\usepackage{hchang}
\usepackage{cleveref}
\usepackage{comment}
\usepackage{thmtools}
\usepackage{thm-restate}
\usepackage{titling}

\usepackage{graphicx} 

\usepackage{amsmath}
\usepackage{setspace}
\usepackage{enumerate}
\usepackage[parfill]{parskip}
\usepackage{comment}
\usepackage{lineno}
\usepackage{xspace}
\usepackage[most]{tcolorbox}
\usepackage[dvipsnames]{xcolor}
\usepackage{fullpage}

\usepackage{titlesec}

\usepackage{caption}

\usepackage{multirow}   

\newtheorem{theorem}{Theorem}
\newtheorem{lemma}{Lemma}
\newtheorem{corollary}{Corollary}

\newcommand{\eps}{\varepsilon}
\newcommand{\cA}{\mathcal{A}}

\newcommand{\cC}{\mathcal{C}}

\newcommand{\cS}{\mathcal{S}}
\newcommand{\cH}{\mathcal{H}}
\DeclareMathOperator{\HD}{HD}
\newcommand{\lifted}{\textsc{lifted}}

\nolinenumbers

\title {Strong Lower Bounds for Weak Epsilon-Nets}

\author{Sujoy Bhore\thanks{Department of Computer Science \& Engineering, Indian Institute of Technology Bombay, Mumbai, India. Email: sujoy@cse.iitb.ac.in. This work is supported in part by ANRF ARG-MATRICS, Grant 002465} \and 
Reilly Browne\thanks{Department of Computer Science, Dartmouth College, New Hampshire, United States. Email: reilly.browne.gr@dartmouth.edu }
\and 
Da Wei Zheng\thanks{Institute of Science and Technology Austria, Klosterneuburg, Austria. Email: dzheng@ista.ac.at. This research was supported by  the Austrian Science Fund (FWF) under grant DOI 10.55776/Z422.
\\For open access purposes, the authors have applied a CC BY public copyright license to any author-accepted manuscript version arising from this submission. }
}

\date{}

\begin{document}

\maketitle

\begin{abstract}
Let $P$ be a finite point set in $\R^d$. A \emph{weak $\eps$-net} for $P$ with respect to convex sets is a finite set $X\subset\R^d$ intersecting every convex set containing at least $\eps|P|$ points of $P$.
Determining their optimal worst-case size was described by Alon, Kalai, Matou\v{s}ek, and Meshulam 
[Adv. Appl. Math., 2002]
as \emph{``one of the truly important open problems in combinatorial geometry.''}

We give elementary constructions which establish the first polynomially superlinear lower bounds in fixed dimension. Already in $\R^3$, we construct point sets for which every weak $\eps$-net has size $\Omega(\eps^{-3/2})$. More generally, for every fixed $d\ge3$, we obtain a lower bound of $\Omega_d(\eps^{-\gamma_d})$, where $\gamma_d=\sqrt{d/2}+O(1)$. Consequently, no upper bound of the form $O_d(\eps^{-C})$ can hold with an absolute constant $C$ which is independent of the dimension. Despite the recent progress on upper bounds, the strongest previous general lower bound remained $\Omega_d(\eps^{-1}\log^{d-1}(1/\eps))$, due to Bukh, Matou\v{s}ek, and Nivasch 
[Israel J. Math., 2011].
To obtain our results, we apply a quadratic lift to grid-based configurations with many point-line incidences.
The bounds also yield polynomial separations between fractional and integral piercing numbers of convex families, and lower bounds for Hadwiger--Debrunner numbers.
\end{abstract}

\section{Introduction}

The theory of $\eps$-nets plays a foundational role in discrete and computational geometry, learning theory, and high-dimensional probability, among other areas; see, e.g.,~\cite{MV17,blum2020foundations,vershynin2019high}. Haussler and Welzl introduced the notions of strong and weak $\eps$-nets in their seminal work on geometric range searching in the 1980s~\cite{HausslerW86,HausslerW87}. Let $\mathcal R$ be a family of subsets of $\R^d$, let $P\subset\R^d$ be a finite nonempty point set, and let $0<\eps\le1$. A \emph{strong $\eps$-net} for $P$ with respect to $\mathcal R$ is a set $X\subseteq P$ that pierces every range $R\in\mathcal R$ containing at least $\eps|P|$ points of $P$. A \emph{weak $\eps$-net} satisfies the same piercing requirement but allows the points of $X$ to lie anywhere in $\R^d$.
We study weak $\eps$-nets for convex ranges. We denote by $\cC_\eps$ the family of all convex sets containing at least $\eps|P|$ points of $P$.

Many families of convex ranges, such as halfspaces or balls in fixed dimension, admit strong $\eps$-nets whose size is independent of $|P|$. This follows from the classical \emph{$\eps$-net theorem}. Every finite range space of Vapnik--Chervonenkis (VC) dimension $\delta$ admits a strong $\eps$-net of size $O((\delta/\eps)\log(1/\eps))$, for $0<\eps\le1/2$~\cite{HausslerW86,HausslerW87,KomlosPW92}. This bound is optimal in general. For every fixed $\delta\ge2$ and sufficiently small $\eps$, there are range spaces of VC dimension at most $\delta$ in which every strong $\eps$-net has size $\Omega((\delta/\eps)\log(1/\eps))$~\cite{KomlosPW92}. Thus, bounded VC dimension guarantees strong $\eps$-nets of size nearly linear in $1/\eps$, and the logarithmic factor cannot be removed in general.

\begin{figure}[ht]
    \centering
    \includegraphics[width=0.5\linewidth,page=1]{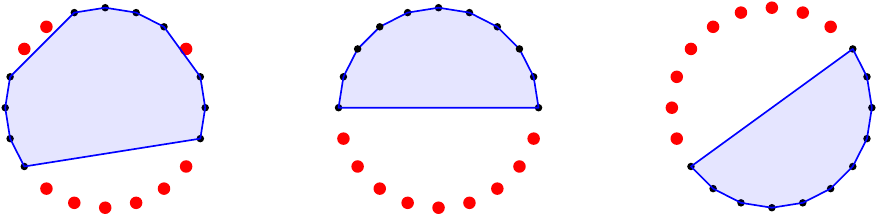}
    \caption{For $n$ points in convex position, any candidate strong $\eps$-net $X$ with $|X|\le n-\lceil\eps n\rceil$ misses the heavy convex set $\conv(P\setminus X)$.}
    \label{fig:strong_hopeless}
\end{figure}

On the other hand, arbitrary convex ranges exhibit a fundamentally different situation. In every dimension $d\ge2$, the family of all convex sets has unbounded VC dimension. For instance, consider a set  $P$ of $n$ points in convex position; then $\conv(S)\cap P=S$ for every subset $S\subseteq P$. Consequently, any set $X\subseteq P$ with at most $n-\lceil\eps n\rceil$ points misses the convex hull of $P\setminus X$, which contains at least $\eps n$ input points; see Figure~\ref{fig:strong_hopeless}. Every strong $\eps$-net for such a configuration therefore requires at least $n-\lceil\eps n\rceil+1$ points. For any fixed $\eps<1$, this rules out a bound independent of $n$.

Allowing net points outside $P$ can dramatically reduce the number of points required, even in the plane. Let $P$ consist of the vertices of a regular $n$-gon centered at the origin, where $n\ge4$ is even, and set $\eps=(n/2+1)/n$. Every convex set containing at least $\eps n$ vertices contains an antipodal pair and hence the origin. Thus, the origin alone forms a weak $\eps$-net, whereas the preceding argument shows that every strong $\eps$-net contains at least $n/2$ points; see Figure~\ref{fig}.

\begin{figure}[ht]
\centering
\includegraphics[width=0.5\linewidth,page=2]{figures.pdf}
\caption{For the vertices of a regular even $n$-gon and $\eps=(n/2+1)/n$, the origin alone forms a weak $\eps$-net, while every strong $\eps$-net requires at least $n/2$ points.}
\label{fig}
\end{figure}

Throughout the paper, we write $r:=1/\eps\ge1$ and discuss weak $\frac1r$-nets. For a finite nonempty set $P\subset\R^d$, let $f_d(P,r)$ denote the minimum cardinality of a weak $\frac1r$-net for $P$ with respect to all convex ranges in $\R^d$. We define the corresponding worst-case size by
\[
f_d(r):=\sup\bigl\{f_d(P,r): P\subset\R^d
\text{ is finite and nonempty}\bigr\}.
\]

A fundamental theorem of Alon, B\'ar\'any, F\"uredi, and Kleitman shows that $f_d(r)$ is bounded by a polynomial in $r$ for every fixed dimension $d$~\cite{AlonBFK92}. Thus, despite the unbounded VC dimension of convex ranges, convexity guarantees weak nets whose size is independent of the number of input points.

Weak $\eps$-nets also provide a fundamental link between fractional and integral piercing of convex sets. A fractional piercing assigns nonnegative weights to finitely many points so that every set in a given convex family receives total weight at least one. If the total weight is $t$, normalization makes every set $\frac1t$-heavy under the resulting probability distribution. A weak net for this distribution then pierces the entire family. Bounds on weak nets therefore control the cost of converting fractional solutions into actual piercing sets~\cite{alon2002transversal}. In fact, this passage is a central step in Alon and Kleitman's proof of the Hadwiger--Debrunner $(p,q)$-theorem, which combines fractional Helly theory with weak nets to deduce bounded piercing numbers from local intersection conditions~\cite{AlonK92}. Reflecting on the importance of this quantitative problem, Alon, Kalai, Matou\v{s}ek, and Meshulam in their seminal work~\cite{alon2002transversal} wrote: 

\begin{quote}
    \emph{``Finding the correct estimates for weak $\eps$-nets is, in our opinion, one of the truly important open problems in combinatorial geometry''.}
\end{quote}

The search for sharp bounds has generated a sustained and long line of work spanning more than three decades. The original theorem of Alon, B\'ar\'any, F\"uredi, and Kleitman gave the planar bound $f_2(r)=O(r^2)$~\cite{AlonBFK92}. In higher dimensions, Chazelle, Edelsbrunner, Grigni, Guibas, Sharir, and Welzl obtained
\[
f_d(r)=O_d\!\left(r^d\log^{c_d}r\right),
\]
where $c_d$ depends only on $d$, in work first presented at STOC 1993~\cite{ChazelleEGGSW93}. Subsequently, Matou\v{s}ek and Wagner gave a simpler construction using simplicial partitions~\cite{Matousek04}. These bounds resisted any improvements in their polynomial exponents for more than two decades.
Rubin finally broke the quadratic barrier in the plane at FOCS 2018 by proving $f_2(r)=O_\eta(r^{3/2+\eta})$ for every fixed $\eta>0$~\cite{Rubin18}. In a subsequent work~\cite{Rubin21} at STOC 2021, Rubin improved the exponent below $d$ in every fixed dimension $d\ge3$. See \Cref{tab:results} for the precise exponents he attained.

These advances substantially reduced the upper bounds, while leaving a large gap from the known lower bounds. Bukh, Matou\v{s}ek, and Nivasch, in their seminal work~\cite{bukh2009lower} (see also~\cite{BukhMN11}) at SoCG 2009, established
\[
f_d(r)=\Omega_d\!\left(r\log^{d-1}r\right)
\]
for every fixed $d\ge2$, which gave the first superlinear lower bound in fixed dimension. 
This remained the strongest general lower bound prior to our work. Although their result ruled out linear-size weak nets, its improvement over the linear bound was only polylogarithmic in every fixed dimension.

The gap therefore left open a basic and fundamental question about the growth of $f_d(r)$. Could weak nets always have size $r^{1+o(1)}$ in every fixed dimension, or does some fixed dimension require polynomially superlinear size? Despite the upper bounds having exponents that grow with $d$, no lower bound of the form $\Omega(r^{1+\delta})$, for a fixed $\delta>0$, was known in any fixed dimension.

\subsection{Our results}

\begin{theorem}
\label{thm:weak-epsilon-nets-general-d}
For every integer $d\ge 3$, define
\[
    \gamma_d
    :=
    \frac12\left\lfloor
        \frac{\sqrt{8d+25}-1}{2}
    \right\rfloor.
\]
There exist constants $r_d>0$ and $c_d$, 
such that for every $r\ge r_d$
there is a point set $P\subset\R^d$ for which every weak $\frac 1r$-net with respect to convex sets has at least $c_dr^{\gamma_d}$ points. Consequently $f_d(r) = \Omega_d\!\left(r^{\gamma_d}\right)$.
\end{theorem}

The following table illustrates the dimensions required to obtain the first few distinct values of $\gamma_d$.
\begin{table}[ht] 
    \centering
\[
\begin{array}{c c c}
    \textbf{Dimension }d
    & \textbf{Lower bound for }f_d(r)
    & \textbf{Upper bound for }f_d(r)\\
    \hline \rule{0pt}{2.4ex}
    2 & \Omega(r\log r) \;\cite{bukh2009lower} & O(r^{1.5+\delta}) \;\cite{Rubin18}\\[1pt]
    \hline
    \begin{array}{c}
        \rule{0pt}{2.4ex}3\\[3pt]
        4\\[3pt]
        5\\[3pt]
        6
    \end{array}
    &
    \textcolor{BrickRed}{\Omega(r^{1.5})}
    &
    \begin{array}{c}
        \rule{0pt}{2.8ex}
        O(r^{2.558+\delta})\;\cite{Natan-arxiv21} \\[3pt]
        O(r^{3.480+\delta})\;\cite{Natan-arxiv21} \\[3pt]
        O(r^{4.437+\delta})\;\cite{Natan-arxiv21}\\[3pt]
        O(r^{5.450+\delta})\;\cite{Natan-arxiv21}
    \end{array}
    \\[3pt]
    \hline
    \rule{0pt}{4.9ex}
    \begin{array}{c}
    7 \\ \vdots \end{array}  & 
    \textcolor{BrickRed}{\Omega(r^2)}     &
    \begin{array}{c}
    O(r^{6.459+\delta})\;\cite{Natan-arxiv21}  \\
    \vdots
    \end{array}
    \\[3pt]
    \hline
    \rule{0pt}{5.1ex}
    \begin{array}{c}
    12 \\
    \vdots
    \end{array} 
    & \textcolor{BrickRed}{\Omega(r^{2.5})} & 
    \begin{array}{c}
    O(r^{11.478+\delta})\;\cite{Natan-arxiv21}\\[3pt]
    \vdots
    \end{array}
\end{array}
\]
    \caption[Caption for results table]{Table of current best lower and upper bounds for $f_d(r)$. New results in this paper are written in red. The existing upper bounds\footnotemark~hold for an arbitrary value $\delta > 0$.}
    \label{tab:results}
\end{table}
\footnotetext{The table uses the corrected full version \cite{Natan-arxiv21} of the 2021 STOC paper \cite{Rubin21}.}

The starting point for our theorem is to apply a quadratic lift $\psi(x,y) = (x,y,x^2, xy)$ to a configuration of points $P_0$ on a 2D grid following Elekes'~\cite{Elekes02} construction with many point-line incidences. Define $P_\lifted =\psi(P_0)$. For every 2D line $\ell$, the lifted points $\psi(\ell\cap P_0)$ are in convex position.
We examine the 2D $q$-rich lines $\ell$ that contain many 2D points of $P_0$, and 4D convex hulls of point sets that contain exactly $q/2$ points of $\psi(\ell \cap P_0)$.
We show that it is difficult to construct a weak net that hits all such 4D convex hulls.
It can be proved that points of a weak net that are not points of $P_\lifted$ can only pierce convex sets originating from a unique line.
We can replace points in the net that are not in the original point set by a small number (determined by the line richness) of lifted points to obtain a strong net. However, a strong net for the lift of points along a line is a strong net for points in convex position, and thus must be large. By a simple counting argument the strong net, and the weak net must also be large. This yields a lower bound of $\Omega(r^{1.5})$ in 4D.

We can reduce the dimensionality of the lower bound to 3D by slightly shifting the grid before performing a modified lift $\phi(x,y) = (x^2,xy, y^2)$. This allows for a point in a weak net to pierce convex sets arising from multiple lines. 
The shift ensures every 4D point outside the lifted grid pierces convex sets arising from at most four distinct 2D lines. 
This generalizes to higher dimensions where we take the quadratic lift of a higher dimensional grid.
To be precise we lift an $s+1$ dimensional grid to $s-1 +\binom{s+2}{2} = (s^2+5s)/2$ dimensions. 

The quadratic lift is essential in our construction since we need to deal with the convex combination of lifted points that originated from a single hyperplane. The crucial idea is that the convex combination allows us to solve for the defining equation of the hyperplane. We are unable to see how to avoid lifting to quadratically many dimensions, as all coordinates of the lift are needed for the hyperplane recovery step.
The idea of using quadratic lifts is not new. A similar idea was used for geometric data structure lower bounds by Erickson \cite[Section 8]{Erickson00}, that employed a 
\emph{quadratic Veronese map} that sends $p=(p_1,\ldots,p_m)$ to
the vector $(p_i p_j)_{1\le i\le j\le m}$ of all pairwise coordinate
products. 
The quadratic Veronese map was used to turn incidence bounds into halfspace relations. 
Afshani~\cite{Afshani13} used a similar map for slabs.

\subsection{Applications}\label{sec:applications}

Our constructions also have consequences for related fundamental problems in discrete geometry. They yield lower bounds for Hadwiger--Debrunner numbers, which quantify the number of piercing points guaranteed by local intersection conditions. Our results also yield polynomial separations between fractional and integral piercing numbers, which establish lower bounds on the integrality gap of the standard linear programming relaxation for minimum piercing set.

\subsubsection{$(p,q)$-Theorem and Hadwiger-Debrunner numbers}
A family of sets $F$ satisfies the $(p,q)$-property if every among any $p$ sets in $F$, there are $q$ sets which share a non-empty intersection. 
It was posed by Hadwiger and Debrunner~\cite{HadwigerD57} whether every family of convex sets in $\R^d$ with $p\ge q \ge d+1$  satisfying the $(p,q)$-property can be pierced by a set of points whose size depends only on $d$, $p$, and $q$. Alon and Kleitman~\cite{AlonK92} prove this in their celebrated $(p,q)$-theorem. This value is known as the \emph{Hadwiger-Debrunner number} and is denoted by $\HD_d(p,q)$. To be precise the Hadwiger–Debrunner number $\HD_d(p,q)$ is the least integer $N$ such that every finite family of nonempty convex sets in $\R^d$, with at least $p$ members and satisfying the $(p,q)$-property, can be pierced by at most $N$ points.

A sequence of work by Keller and Smorodinsky~\cite{KellerS21} and Roche-Newton~\cite{RocheNewton26} use the general position construction of Balogh and Solymosi~\cite{BaloghS18} to show $\HD_2(p,3)\ge \Omega(p^{6/5-\eps})$.
In higher dimensions, Keller and Smorodinsky~\cite{KellerS26} utilized the higher dimensional general position constructions of Suk and Zeng~\cite{SukZ26} that show that for $d\ge 3$, that $\HD_d(p,d+1)\ge \Omega(p^{2-\eps})$ for any $\eps > 0$.

Lower bounds for weak $\varepsilon$-nets transfer directly to Hadwiger--Debrunner numbers~\cite{GovindarajanN15,KellerS21}. Indeed, given a finite point set $P$, let $\cC$ 
be the family of convex hulls of subsets of $P$ containing at least $(q/p)|P|$ points. Among any $p$ members of $\cC$, a simple counting argument shows that there must be a point of $P$ contained in at least $q$ of them. Thus this family satisfies the $(p,q)$-property, while every piercing set is a weak $(q/p)$-net with respect to all convex sets. Consequently,
$\HD_d(p,q)\ge f_d(p/q)$ so a lower bound of the form $f_d(r) = \Omega(g(r))$ implies a lower bound of $\HD_d(p,q) = \Omega(g(\frac pq))$.
The weak net lower bound due to Bukh, Matou\v{s}ek, and Nivasch~\cite{BukhMN11} results in a lower bound of $\HD_d(p,q) = \Omega_d(\frac{p}{q}\log^{d-1}(\frac{p}{q}))$. 
A corollary of our result is a new lower bound for $\HD_d(p,q)$.
\begin{corollary}
For any dimension $d\ge 3$, and all $p\ge q\ge d+1$,
    \[\HD_d(p,q) = \Omega_d\left(\left(\frac{p}{q}\right)^{\gamma_d}\right) 
    \qquad \text{where } \qquad
    \gamma_d
    =
    \frac12\left\lfloor
        \frac{\sqrt{8d+25}-1}{2}
    \right\rfloor.
    \]
\end{corollary}
We observe that this leads to two refinements of existing lower bounds on $\HD_d(p,q)$. The first is that this yields a lower bound where the exponent increases beyond $2$ provided $d\ge 12$. The second is that even if $q = p^\alpha$ for some constant $0<\alpha <1$, in sufficiently high dimension $d$, the quantity $\HD_d(p,q)$ is superlinear in $p$.

\subsubsection{Approximating Minimum Piercing Set}

For a set system $F = (U, \cS)$, the minimum hitting set is the minimum cardinality set $H \subset U$ such that for every $S \in \cS$, $S \cap H \neq \emptyset$. In this sense, a strong $\eps$-net is a hitting set for the set system $(U, \cC_\eps)$ and a weak $\eps$-net is a hitting set for $(\mathbb{R}^d, \cC_\eps)$. Let $\tau(F)$ be the cardinality of the minimum hitting set of $F$.
When $U = \R^d$, we will use the term \textit{minimum piercing set}.

On the positive side, the existence of $g(\frac{1}{\eps})$-sized strong $\eps$-nets for a given set system implies a polynomial time algorithm which returns a hitting set of size $g(4\tau(F))$, due to Br\"{o}nnimann and Goodrich \cite{Bronnimann1995}. This algorithm relies on the multiplicative weight update paradigm, wherein the weights of a finite set of points $P$ are modified when the calling to the $\eps$-net construction fails to return a valid hitting set.
In the case of approximating a minimum piercing set, all points in $\R^d$ are valid elements to use for hitting and the sets to be hit are the input convex regions $\cC_\eps$. To make the universe finite, assign a representative point from each cell in the arrangement $\cA$ induced by $\cC_\eps$. Weak $\eps$-nets in $(\R^d, \cC_\eps)$ therefore correspond to strong $\eps$-nets in the set system $(\cA, \cC_\eps)$. Thus, Rubin's \cite{Natan-arxiv21} upper bound implies polynomial time approximation algorithms that are at worst polynomial in the optimal solution, namely $\tau(F)^{O(d)}$ for fixed $d$.

The existence of weak $\eps$-nets for $P$ with respect to $\cC$ is also intimately related to the linear programming relaxation of minimum piercing set. Here, we use the standard relaxation for hitting set in system $(\cA, \cC)$:
\begin{align*}
    \text{minimize} & \sum_{p \in \cA} x_p \\
    \text{s.t.} & \sum_{p \in C} x_p \geq 1 \hspace{1em} \forall C \in \cC \\
    & 0 \leq x_p \leq 1 \hspace{1em} \forall p \in \cA
\end{align*}

Let $\tau^*(F)$ be the weight of the solution to the corresponding linear programming relaxation, also called the \emph{fractional piercing number}. For a point set $P$ weighted in accordance with the solution to the LP, every convex set in $\cC$ has weight at least $1$, and thus a $\frac{1}{\tau^*(F)}$-net gives a piercing set for $\cC$. Conversely, assigning weight $\frac{1}{\eps|P|}$ to all points in $P$ yields a fractional solution to the linear programming relaxation for $(\R, \cC_\eps)$. This means that a lower bound on the size of an $\eps$-net gives a lower bound for the integrality gap ratio $\frac{\tau(F)}{\tau^*(F)}$. Our improved lower bound yields:

\begin{corollary} For every dimension $d\ge 3$ there is a constant $c_d$ such that for every sufficiently large $r$, there exists a finite family $F$ of convex sets in $\R^d$ such that:

\[\tau^*(F) \le r \quad \text{ and }
\quad
\tau(F) \ge c_dr^{\gamma_d}
    \qquad \text{where } \qquad
    \gamma_d
    =
    \frac12\left\lfloor
        \frac{\sqrt{8d+25}-1}{2}
    \right\rfloor,
\]
giving an integrality gap of $\frac{\tau(F)}{\tau^*(F)} =\Omega((\tau^*(F))^{\gamma_d-1})$.
\end{corollary}

\subsection{Organization}
We begin in Section~\ref{sec:2dlift} by presenting two different lifts of the 2D grid into higher dimension. The first is a simple lifting construction into 4D which illustrates the key ideas in \Cref{ssec:4D}. Using these ideas, we can strengthen the result by using a modified lift of a slightly shifted grid into just 3 dimensions in \Cref{ssec:3D}. From there, we proceed to generalize the construction in \Cref{sec:higherD}.

\section{Lifting a 2D Grid for Lower Bounds} \label{sec:2dlift}
\subsection{Lower Bound in Four Dimensions} \label{ssec:4D}
In this section we describe a $\Omega(r^{3/2})$ lower bound for $\frac 1r$-nets in 4D.
Our starting point is the Elekes grid construction~\cite{Elekes02} of point-line incidences in the plane that shows the Szemer\'edi--Trotter incidence bound is tight \cite{SzemerediT83}.
Consider a grid of points $P_0$ and a set of lines $L$ defined as follows for an even integer $q\ge 2$.
\[ P_0 := [q]\times[2q^2] \qquad \text{and} \qquad L = \{y=ax+b:a\in[q], b\in[q^2]\} \]
It is easy to verify that $|P_0| = 2q^3$, $|L| = q^3$, each line contains exactly $q$ points, and each point lies on at most $q$ lines.

We lift the point set to $\R^4$ using the injective map $\psi:\R^2\to \R^4$ defined by $\psi(x,y) = (x,y,x^2, xy)$. Observe that for any line $\ell\in L$, the lift $\psi$ maps the points of the line onto a quadratic curve, and thus all points of $\phi(\ell\cap P_0)$ lie in convex position.

We will consider the points $P := \psi(P_0) =  \{\psi(p):p\in P_0\} $ and the following family of convex sets $\cC_\ell$ in $\R^4$ each with at least $q/2$ points on a line $\ell$.
\[  \cC_\ell = \{C=\conv(\psi(C_0)): C_0 \text{ is a set of exactly $q/2$ points of $\ell\cap P_0$} \}  \qquad \cC = \bigcup_{\ell\in L} \cC_\ell\]
Each set of $\cC$ contains at least $q/2$ points out of the $2q^3$ total number of points. In the remainder of this section suppose $X$ is a weak $\frac1r$-net with $r=4q^2$ for $P$. Observe that $X$ must hit each set of $\cC$. 

Consider an arbitrary $z= (z_1,z_2,z_3,z_4) \in X$, that hits some set $\conv(\psi(C_0))\in \cC_\ell$, where $C_0$ contains exactly $q/2$ points $\{(x_1,y_1), \dots, (x_{q/2},y_{q/2})\}$ of $\ell\in L$. Let $\ell$ be defined by the equation $y=ax+b$.
We show that either $z$ is a point of the form $\psi(x_i,y_i)$, or $z$ can pierce sets only from the family $\cC_\ell$.

Since $z$ is in the convex hull, there must be some nonnegative coefficients $\lambda_i$ where $\sum_{i=1}^{q/2}\lambda_i = 1$  and that $z = \sum_{i=1}^{q/2}\lambda_i \psi(x_i,y_i)$.
It holds that 
\begin{equation}\label{eq:4d_1}
    z_2 = \sum_{i=1}^{q/2}\lambda_i y_i = \sum_{i=1}^{q/2}\lambda_i(ax_i + b) = az_1 + b.
\end{equation} 
Furthermore, as $y_i=ax_i+b$ for all $i$,
\begin{equation}\label{eq:4d_2}
z_4 = \sum_{i=1}^{q/2}\lambda_i x_iy_i = \sum_{i=1}^{q/2}\lambda_i(ax_i^2 + bx_i) = a\sum_{i=1}^{q/2}\lambda_i x_i^2 +b\sum_{i=1}^{q/2}\lambda_i x_i =  az_3 + bz_1.
\end{equation}
Note that these two equations define relations on $z_i$ that holds regardless of the choice of convex representations $\lambda_i$. Eliminate $b$ by subtracting $z_1$ times \Cref{eq:4d_1} from \Cref{eq:4d_2}. This yields
\begin{equation*} 
z_4 - z_1z_2 = a(z_3 - z_1^2).
\end{equation*}
Observe that if $z_3 - z_1^2\neq 0$, we can uniquely recover $a$ from the coordinates of $z$ by dividing $(z_3-z_1^2)$ on both sides, and also uniquely recover $b$ from \Cref{eq:4d_1}. Otherwise, it remains to consider $z_3-z_1^2 = 0$, where we have that
\[
    0= z_3-z_1^2 = \sum_i\lambda_i x_i^2 - \left(\sum_i\lambda_i x_i\right)^2
    =
    \sum_i\lambda_i(x_i-z_1)^2.
\]
Thus every $x_i$ with $\lambda_i>0$ has $x_i = z_1$. Since the points $(x_i, y_i)$ lie on a non-vertical line $\ell$, we conclude that $z$ is the lift of a single point so $z\in P$.

\paragraph*{Size of the weak net} Now we can bound the size of the weak $\frac1r$-net $X$.
To do so,
we will replace points of $X$ by lifted grid points while preserving the sets of $\cC$ that are pierced.
Consider any point $z\in X\setminus P$. We have shown that $z$ can only pierce convex sets from $\cC_\ell$ for at most one line $\ell_z \in L$. Thus if we replace $z$ with the points $Q_z:= \{\psi(x):x\in P_0\cap \ell_z\}$, any convex set of $\cC_{\ell_z}$ that was pierced by $z$ is also pierced by $Q_z$. Define $Q:=\bigcup_{z\in X\setminus P} Q_z$. Now the set $X' := (X\cap P) \cup Q$ must contain a strong $\frac12$-net for $P_\ell := \psi(\ell\cap P_0)$ for every $\ell \in L$ as $X'$ pierces all sets of $\cC_\ell$ that have exactly $q/2$ points. By the same argument as in the introduction, such a strong net must have at least $q/2+1$ points of $P_\ell$. 
As a single point of $X\cap P$ belongs to at most $q$ sets $P_\ell$, and each $Q_z$ has at most $q$ points and contributes to at most one line $\ell_z$,  
\[ q^3\cdot (q/2+1) \le \sum_{\ell\in L} |X\cap P_\ell| + \sum_{z\in X\setminus P}|Q_z| \le   q|X\cap P| + q|X\setminus P| \qquad \Rightarrow \qquad |X| \ge q^3/2 + q^2.\]
This proves the following result.

\begin{theorem}
For every even integer $q\ge 2$ there is a set $P\subset \R^4$ of $2q^3$ points whose weak $\frac1{4q^2}$-nets have size at least $q^3/2+q^2$. Consequently, $f_4(r) = \Omega(r^{3/2})$.
\end{theorem}

\subsection{Improved Lift to Three Dimensions}\label{ssec:3D}
We use the quadratic lift $\phi: \R^2 \to \R^3$ defined by $ \phi(x,y) = (x^2, xy, y^2)$. We will lift the same grid construction as before with a parameter $q\ge 2$ (assumed to be even for convenience), but with a crucial difference. Before applying the lift, translate the entire grid upwards by $\tau = \frac{1}{9q^5}$. To be precise, we define the following:
\[ P^{+\tau} = \{(a,b+\tau): a\in [q], b\in [2q^2]\} \qquad \text{and} \qquad L^{+\tau} = \{y=ax+b+\tau: a\in [q], b\in [q^2]\}.\]
Define as before $P = \phi(P^{+\tau})$ and $\cC_\ell$ as well as $\cC$ as follows:
\[
    \cC_\ell:=
    \{\conv(\phi(S)):
        S\subseteq P^{+\tau}\cap\ell,\ |S|=q/2\},
    \qquad
    \cC:=\bigcup_{\ell\in L^{+\tau}}\cC_\ell.
\]

\begin{figure}[t]
    \centering
    \includegraphics[width=0.99\linewidth]{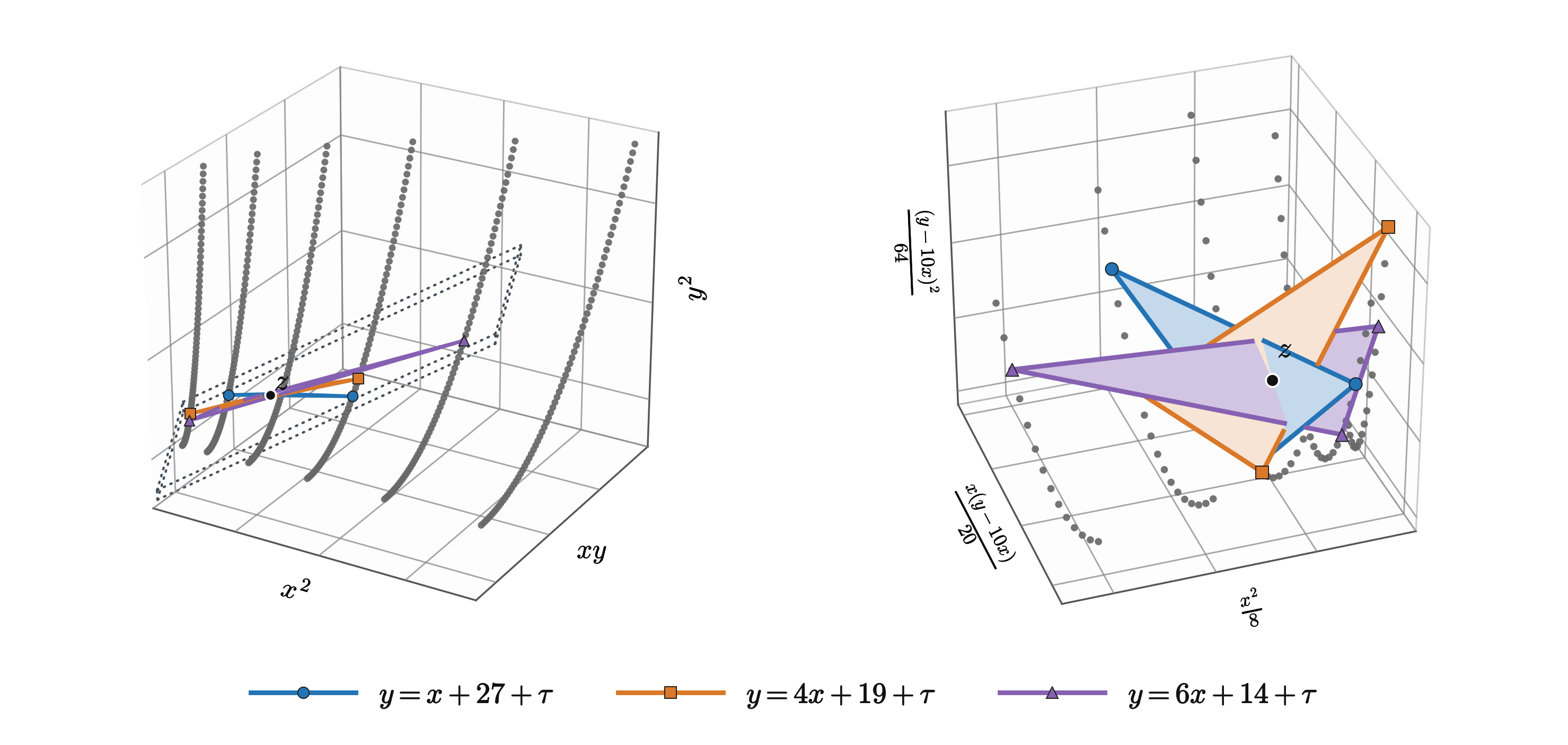}
    \caption{On the left is a plot of the lifted points for $q=6$ and $\tau = 1/69984$ along with three intersecting triangles arising from three distinct lines. These triangles intersect at a point $z$. The dotted box on the left denotes the skewed and rescaled view on the right where it is more clear that the three triangles are intersecting at $z=\left(
\frac{241}{30}+\frac{\tau}{15},\;
\frac{977}{12}+\frac{17\tau}{6},\;
\frac{4419}{5}+\frac{298\tau}{5}+\tau^2
\right)$. }
    \label{fig:3Dprojections}
\end{figure}
Let $X$ be a weak $\frac1r$-net for $P$, and consider a point $z=(z_1,z_2,z_3)\in X$ that hits the convex set of the lift of $q/2$ points $\{(x_1,y_1),\dots,(x_{q/2},y_{q/2})\}$ of some line $\ell\in L^{+\tau}$.
We now show that either $z$ is a point of $P$, or that $z$ pierces sets of $\cC_{\ell_i}$
for at most four distinct lines $\ell_1,\ell_2,\ell_3,\ell_4\in L^{+\tau}$.

Suppose $z$ hits a set in $\cC_\ell$ for $\ell\in L^{+\tau}$ that is defined by $y=ax+b+\tau$ for which we can write $z=\sum_{i}\lambda_i\phi(x_i,y_i)$ with $(x_i, y_i)$ on $\ell$, and $\lambda_i > 0$  with $\sum_i \lambda_i =1$. Thus we can define a quadratic function that depends only on $z$ as follows:
\begin{align*}
    G_z(t) = \sum_i \lambda_i (tx_i -y_i)^2  = t^2\sum_i \lambda_i x^2_i-2t\sum_i \lambda_ix_iy_i + \sum_i \lambda_i y_i^2 = z_1t^2 - 2z_2t + z_3.
\end{align*}
Furthermore, evaluating $t$ at the slope of $\ell$ yields 
\[ G_z(a) = \sum_i \lambda_i (ax_i -y_i)^2 = (b+\tau)^2 \]
Now we will use the following lemma about $G_z$ that we will defer to later.
\begin{lemma} \label{lem:3D-4sol}
If a real polynomial $G(t)$ of degree two satisfies $G(a) = (b+\tau)^2$  for at least five distinct pairs $(a,b)\in [q]\times [q^2]$, then $G(t) = \alpha(t-t_0)^2$ for some $\alpha,t_0\in \R$.
\end{lemma}
Suppose that $z$ hits sets from at least five distinct
line families.
Then, \Cref{lem:3D-4sol} gives $z_1t^2-2z_2t+z_3 = G_z(t) = \alpha(t-t_0)^2$. Observe that as $z_1 = \sum_i \lambda_i x_i^2 >0$, we must have $\alpha >0$. Evaluating at $t_0$ gives 
\[ 0 = G_z(t_0) = \sum_{i} \lambda_i (t_0x_i-y_i)^2. \]
This can only happen if $t_0x_i = y_i$ for all $i$ with $\lambda_i >0$. 

Since also $y_i=ax_i+b+\tau$, we obtain $(t_0-a)x_i=b+\tau>0$.
Thus $t_0\ne a$, and all represented points have the
same $x$-coordinate and hence the same $y$-coordinate.
Consequently, $z$ is the lift of a single grid point,
so $z\in P$.

Thus we have shown that either $z\in P$ or that $z$ can hit convex sets from at most four lines $\ell\in L^{+\tau}$. 
Recalling the prior argument, each point in $X\cap P$ is a part of at most $q$ lines, and each point of $X\setminus P$ can be replaced by $q$ points on $4$ different lines.  This gives 
\[ q^3\left(\frac{q}2+1\right)\le q|X\cap P| + 4q|X\setminus P| \le 4q|X| \qquad \Rightarrow\qquad  |X| \ge q^3/8+q^2/4. \]
This proves the following 3D lower bound.
\begin{theorem}\label{thm:3D}
For every even integer $q\ge 2$, there is a set $P\subset \R^3$ of $2q^3$ points whose weak $\frac1{4q^2}$-nets have size at least $q^3/8+q^2/4$. Consequently, $f_3(r) = \Omega(r^{3/2})$.
\end{theorem}

It remains to prove \Cref{lem:3D-4sol}, that $G_z$ must either be a square of a linear polynomial or has at most four distinct positive integer solutions.
\begin{proof}[of \Cref{lem:3D-4sol}]
Consider five unique integer solutions $G(a_i)=(b_i+\tau)^2$ with $a_i\in [q]$ and $b_i\in [q^2]$ for $i\in [5]$. 
Note that the slopes $a_i$ are distinct since $b$ is uniquely determined by $a$.
Consider if we instead interpolated at the first three solutions polynomials of degree at most two satisfying
\[ R(a_i) = b_i, \qquad T(a_i) = b_i^2, \quad \text{ for } \quad i=1,2,3.\]
Lagrange interpolation gives 
\[ 
R(t) = \sum_{i=1}^3 b_i \prod_{\substack{j=1\\j\neq i}}^3 \frac{t-a_j}{a_i-a_j}
\qquad \text{and} \qquad 
T(t) = \sum_{i=1}^3 b_i^2 \prod_{\substack{j=1\\j\neq i}}^3 \frac{t-a_j}{a_i-a_j}.
\]
Without loss of generality, we may assume that $a_1 < a_2 < a_3$. Observe that by setting $N=(a_2-a_1)(a_3-a_1)(a_3-a_2) > 0$, we can clear the denominators so $NR(t)$ and $NT(t)$ are polynomials with integer coefficients.

At these first three points, we have
\[ G(a_i) = (b_i + \tau)^2 = T(a_i) + 2\tau R(a_i) + \tau^2 \]
Since both $G$ and $T+2\tau R+\tau^2$ have degree at most two and agree at three points, they are identical as polynomials, so $G(t) = T(t) + 2\tau R(t) + \tau^2$.
We next show that $R(a_i) = b_i$ for $i=4,5$. 
For any solution $(a,b)$, 
\begin{align}
(b+\tau)^2 &= G(a) = T(a) + 2\tau R(a) + \tau^2 \nonumber \\
 b^2 + 2\tau b &= T(a) + 2\tau R(a) \nonumber \\
 N(b^2 - T(a)) &= 2\tau N(R(a) -b) \label{eq:integer_polynomial}
\end{align}
The left side of \Cref{eq:integer_polynomial} is clearly integral because $NT(a)$ and $Nb^2$ are integers. 
On the right side, we can bound the absolute value of $2\tau N(R(a) -b)$ for any $a\in [q]$ and $b\in [q^2]$ by using $|N|\le q^3$ and $|NR(a)| \le 3q^5$.
\[  2\tau(|NR(a)| + |Nb|) \le 2\tau(3q^5 +q^5) \le \frac{8q^5}{9q^5} < 1\] 
Since the only integer with absolute value less than one is zero, we conclude that both sides of \Cref{eq:integer_polynomial} vanish for each solution $(a,b)$, so all five solutions satisfy $R(a) = b$. Thus $G(t) - (R(t) + \tau)^2$ has five distinct roots $a_1, \dots, a_5$, but has degree at most four, so must be identically zero. This means $G(t) = (R(t)+\tau)^2$. Since $G$ has degree two, $R$ must have degree one, giving the representation as desired.
\end{proof}
\paragraph*{Remark} Further analysis can show that every point $z\not\in P$ can hit convex sets arising from at most $3$ distinct lines. The bound of $4$ suffices for our purposes, so we omit the lengthier analysis. See \Cref{fig:3Dprojections} for a concrete example of $3$ convex hulls with a common point.

\section{Generalization to Higher Dimensions}\label{sec:higherD}

In this section we give a higher dimensional generalization of our 3D lift.
Fix an integer $s\ge 1$, and an even integer $q\ge 2$. Set $D_s = (s^2+5s)/2$ and $\tau = \frac{1}{9q^5}$ as before. We will construct a point set in $\R^{D_s}$ 
from the shifted grid in $\R^{s+1}$ that supports many rich hyperplanes. Define points $p=(x_1, \dots, x_s, x_{s+1})$ in the shifted grid, with the shifted hyperplanes as below.
\begin{align*} P_0^{+\tau}:= \{(x_1, \dots, x_s, x_{s+1}+\tau): x\in[q]^s, x_{s+1}\in [(s+1)q^2]\} \text{  and  } \\
\cH^{+\tau} = \{H_{a,b}: x_{s+1}=a_1x_1 + \cdots +a_sx_s + b + \tau : a\in [q]^s, b\in [q^2]\}
\end{align*}
Observe that $|P_0^{+\tau}| = (s+1)q^{s+2}$ and $|\cH^{+\tau}| = q^{s+2}$. Let $B_H:=P_0^{+\tau}\cap H$ for $H\in \cH^{+\tau}$. Every choice of $x\in [q]^s$ gives one point of $B_{H}$ so $|B_H| = q^{s}$.
We define the lift $\phi_s:\R^{s+1}\to \R^{D_s}$ as follows:
\[ \phi_s(p):=(x_1, \dots, x_{s-1}, (x_ix_j)_{1\le i\le j\le s+1}) \]
Note that we retain all quadratic coordinates and omit the $x_s$ and $x_{s+1}$ term in the lift. 
Define as before $P^{+\tau} = \phi_s(P^{+\tau}_0)$ and $\cC_H$ to be the family of convex hulls of lifted points with exactly $q^s/2$ points on the hyperplane $H \in \cH^{+\tau}$, and let $\cC = \bigcup_{H\in \cH^{+\tau}} \cC_H$.  
Every member of $\cC$ contains at least $q^s/2$ of the
$(s+1)q^{s+2}$ input points. Consequently, every weak $1/r$-net
must hit every member of $\cC$, where $r = 2(s+1)q^2$.

To generalize the counting argument of the three dimensional construction,
we show that every point of a weak net can be replaced by $O(q^s)$ points of $P^{+\tau}$. 
We defer the following proof to after the counting argument.
\begin{lemma}[Replacement Set Lemma]\label{lem:replacement}
For every point $z\in \R^{D_s}$, 
there exists a collection $Q_{z,H}\subseteq \phi_s(B_H)$ for every $H\in \cH^{+\tau}$ 
such that $z$ pierces $C\in \cC_H$ means that $Q_{z,H} \cap C \neq \varnothing$ and $\sum_{H\in \cH^{+\tau}} |Q_{z,H}| \le 4q^s$.
\end{lemma}

Let $X$ be a weak $\frac1r$-net for $P^{+\tau}$ with $r = 2(s+1)q^2$. For each hyperplane $H\in \cH^{+\tau}$, define $U_H = \bigcup_{z\in X} Q_{z, H}\subseteq \phi_s(B_H)$. By \Cref{lem:replacement}, $U_H$ is a strong $\frac12$-net for $\phi_s(B_H)$. Since the points of $\phi_s(B_H)$ are in convex position, any strong $\frac12$-net must have size at least $q^s/2+1$. Summing over all $q^{s+2}$ hyperplanes of $\cH^{+\tau}$ yields:
\[ q^{s+2}\left(\frac{q^s}{2}+1\right) \le \sum_{H\in \cH^{+\tau}}|U_H| \le \sum_{z\in X}\sum_{H\in \cH^{+\tau}}|Q_{z,H}| \le 4q^s |X| 
\quad \Rightarrow \quad
|X| \ge \frac{q^{s+2}}{8} + \frac{q^2}{4} = \Omega(q^{s+2}) = \Omega(r^{(s+2)/2}).
\]
Thus we have proven the following theorem that is equivalent to \Cref{thm:weak-epsilon-nets-general-d}.
\begin{theorem}
For every $s\ge 1$, there are finite point sets in $\R^{D_s}$ with $D_s= (s^2+5s)/2$ whose weak $\frac{1}{r}$-nets require $\Omega(r^{(s+2)/2})$ points.
\end{theorem}

It remains to prove \Cref{lem:replacement}.
\begin{proof}[of \Cref{lem:replacement}]
Call a set $H\in \cH^{+\tau}$ \emph{compatible} with a point $z$ if $z\in\conv(\phi_s(B_H))$. If no such hyperplane exists, the point is irrelevant and set $Q_{z,H}=\varnothing$ for all hyperplanes $H$. Henceforth we may assume at least one such $H$ exists for $z$.
Fix a representation of $z$ as the convex combination of lifted points $p^{(i)}$ on $H\cap P^{+\tau}_0$.
\[
    z=\sum\lambda_i\phi_s(p^{(i)}),
    \qquad
    p^{(i)}=(p_1^{(i)},\ldots,p_s^{(i)},p_{s+1}^{(i)}),
    \qquad
    \quad \sum\lambda_i=1,
    \qquad
    \lambda_i >0
\]
For notational convenience, write the coordinates of $z$ as:
\[ z = (z_1, \dots, z_{s-1}, (z_{i,j})_{1\le i\le j\le s+1}) \qquad \text{and} \qquad z_{j,i} = z_{i, j} 
.\]

Consider the subspace $F_z:=\operatorname{span}\{p^{(1)}, \dots, p^{(m)}\}$ and let $k:=\dim F_z-1$. We claim that $F_z$ does not depend on the choice of representation for $z$, since for any vector $u\in \R^{s+1}$, we have that:
\[
\sum_{i=1}^{m}\lambda_i(u\cdot p^{(i)})^2
=
\sum_{j=1}^{s+1}\sum_{k=1}^{s+1}u_j u_k \sum_{i=1}^m \lambda_i p^{(i)}_jp^{(i)}_k   
=
\sum_{j=1}^{s+1}\sum_{k=1}^{s+1}u_j u_k z_{j,k}.
\]
Thus $u$ is orthogonal to every $p^{(i)}$ and thus to $F_z$ if and only if the left side is $0$. This can be computed with the coordinates of $u$ and $z$, and is thus independent of the choice of convex combination.

\paragraph*{Replacement points} We define the replacement points $Q_{z,H}:=\phi_s(P_0^{+\tau}\cap F_z\cap H)$ for all $H$ where $z\in \conv(\phi_s(H\cap P_0^{+\tau}))$ and empty otherwise. The set $F_z\cap H$ is a $k$-dimensional flat as $F_z$ contains the origin, but $H$ does not. It is easy to see that the number of grid points on a projected flat has size at most $|Q_{z,H}|\le q^k$, and that the number of hyperplanes that contain the $k$-dimensional flat $F_z \cap H$ is at most $q^{s-k}$. It suffices to show that there are at most four such flats $F_z\cap H$, which would imply that $\sum_{H\in \cH^{+\tau}} |Q_{z,H}| \le 4q^k\cdot q^{s-k} = 4q^s$.

\paragraph*{Hyperplanes containing the flat}
Fix the hyperplane $H=H_{a,b}$ defined by $x_{s+1}=a_1x_1+\cdots + a_sx_s+b+\tau$. 
We know that all points $p^{(i)}$ for $i\in [m]$ lie on $H$.
Inspecting the coordinates of $z$ for $j\in [s-1]$ yields:
\begin{align*}
z_{j,s+1} = \sum_{i=1}^m  \lambda_i p^{(i)}_{j}p^{(i)}_{s+1} = \sum_{i=1}^m \lambda_i p_j^{(i)}\left(b+\tau +\sum_{k=1}^sa_kp^{(i)}_k\right) = (b+\tau)z_j + \sum_{k=1}^s a_kz_{j,k}.
\end{align*}
Observe that if we treat $b$ and $a_k$ as variables and $z$ as fixed, we obtain $s-1$ equations over $s+1$ unknowns. Thus using Gaussian elimination setting additional free variables to $0$, we can express every $a_k$ for $k\in [s-1]$ as a degree at most one equation of (the variables) $a_s$ and $b$.  Any setting of the free variables yields equivalent affine forms and thus the same flat.
Each pair $(a_s,b)$ determines the flat $F_z\cap H$,
although different pairs may determine the same flat.
Let $\hat{a}_k(t, h)$ denote this affine function, and define $\hat{a}_s(t, h) = t$. Note that we can always write $\hat{a}_j(t, h) = \hat a_j(t, 0) + \eta_jh$, where $\eta_j$ is the coefficient of $h$.

For $p\in F_z$, membership in $H$ is equivalent to the
following equation.
\[ 
p_{s+1} 
= \sum_{j=1}^{s-1} (\hat a_j(a_s,0) +(b+\tau)\eta_j)p_j +a_sp_s + b + \tau 
= \sum_{j=1}^s \hat a_j(a_s,0)p_j + (b+\tau)\left(1+\sum_{j=1}^{s-1}\eta_j p_j\right)
\]
Rearranging this equation yields:
\begin{equation} \label{eq:hyperplane}
{p_{s+1}-\sum_{j=1}^{s}\hat a_j(a_s,0)p_j}
=
\left({b+\tau}\right)
\left(
1+\sum_{j=1}^{s-1}\eta_jp_j
\right).
\end{equation}

For fixed $p$ and $z$, the two sides depend on $a_s$ and $b$ respectively.
The LHS motivates the following definition of a quadratic polynomial $ G_z(t)$:
\begin{align*}
{G}_z(t) 
&:= \sum_{i=1}^m \lambda_i\left(p_{s+1}^{(i)} - \sum_{j=1}^s \hat{a}_j(t,0)p_j^{(i)}\right)^2 
\\
&= z_{s+1,s+1} - 2\sum_{j=1}^s \hat{a}_j(t,0) z_{j,s+1} + \sum_{j=1}^s\sum_{\ell=1}^s \hat{a}_j(t,0)\hat{a}_\ell(t,0) z_{j,\ell}.
\end{align*}

By \Cref{eq:hyperplane}, any valid pair $(a_s, b)$ satisfies:
\begin{equation}\label{eq:eval_at_as}
{G}_z(a_s)=(b+\tau)^2
\underbrace{\sum_{i=1}^{m}\lambda_i
\left(1+\sum_{j=1}^{s-1}\eta_jp_j^{(i)}\right)^2}_{\text{determined only by }z}.
\end{equation}
For shorthand, define $\Delta_z$ to be the part that depends only on $z$, so ${G}_z(a_s) = \Delta_z (b+\tau)^2$.

\paragraph*{Counting the flats}
 
By the definition of $G_z$ as a sum of squares, $\Delta_z\ge0$.

Suppose that $\Delta_z=0$, and fix a compatible hyperplane
with last slope $a_s$. Then
\[
    0=G_z(a_s)
     =\sum_i\lambda_i
       \left(
           p_{s+1}^{(i)}
           -\sum_{j=1}^{s}\hat a_j(a_s,0)p_j^{(i)}
       \right)^2.
\]
Discarding zero weights, we may assume that every
$\lambda_i>0$. Thus every represented point satisfies
\[
    p_{s+1}^{(i)}
    =\sum_{j=1}^{s}\hat a_j(a_s,0)p_{j}^{(i)}.
\]
Since this equation is homogeneous and linear in $p$, it holds throughout $F_z$.
Consequently the defining equation of every compatible hyperplane restricted to $F_z$ reduces to $1+\sum_{j<s}\eta_jp_j=0$. Thus all compatible hyperplanes determine the same flat in $F_z$.

We now consider when $\Delta_z>0$.
We break into two cases. In the first case, consider if $ G_z$ is not a square of a degree one polynomial. A nonnegative quadratic that is not a square has no real root, so $\Delta_z \ge 0$. Now we can apply \Cref{lem:3D-4sol} to $G_z/\Delta_z$, and obtain at most four distinct integer solution pairs $(a_s, b)$, and thus at most four distinct flats.

If $G_z$ is constant, the vanishing $t^2$-coefficient in its expression as a sum of squares forces
$p_{s+1}-\sum_{j=1}^s\widehat a_j(t,0)p_j$ to be independent of $t$, hence on its linear span $F_z$.
Since $G_z(a_s)=\Delta_z(b+\tau)^2$ also fixes $b+\tau>0$ when $\Delta_z>0$, all compatible hyperplanes induce 
the same flat in $F_z$.

Otherwise, we can assume that $G_z$ is nonconstant and thus a square so that $G_z(t) = \alpha (t - t_0)^2$ for real values $\alpha,t_0\in \R$. Since also $G_z(a_s) = \Delta_z(b+\tau)^2$, the following ratio takes on at most two distinct values.
\[
\frac{a_s-t_0}{b+\tau}
=
\pm\sqrt{\frac{\Delta_z}{\alpha}}.
\]

As $G_z(t_0)=0$ and $G_z$ is a sum of squares, it follows that for all $p^{(i)}$, the term $t-t_0$ is a factor of $p^{(i)}_{s+1}-\sum_{j=1}^s \hat{a}_j(t, 0)p_j$. By linearity this also holds for all $p\in F_z$. Using the defining hyperplane equation of \Cref{eq:hyperplane}, the LHS of the equation (treating $a_s$ as a variable) is a linear polynomial in $a_s$ with a root at $t_0$, thus can be rewritten as $\alpha'(a_s-t_0) = \beta' (b+\tau)$, where $\alpha'$ and $\beta'$ are constants that depend on $z$ and $p$. Thus the hyperplane equation is invariant up to the value of $\frac{a_s-t_0}{b+\tau}$. Since this can take on at most two values, there are at most two distinct flats.

Over all cases we have shown that there are at most four distinct flats. Recalling that each flat contains $q^k$ grid points, and is contained in at most $q^{s-k}$ hyperplanes, we conclude that the number of replacement points is at most $4q^s$.
\end{proof}

\section*{AI Acknowledgment}
An initial proof of the lower bound result was obtained through discussions with ChatGPT-6 Astra. The authors subsequently verified its correctness, simplified the argument, and substantially revised the exposition. The authors take full responsibility for the results and proofs presented in this paper.

\bibliographystyle{alphaurl}
\bibliography{ref}

\end{document}